# Multi-frame Interferometric Imaging with a Femtosecond Stroboscopic Pulse Train for Observing Irreversible Phenomena

Dmitro Martynowych[1,2,a)], David Veysset[2], A. A. Maznev[1], Yuchen Sun[1,2], Steven E. Kooi[2], Keith A. Nelson[1,2]

[1]Department of Chemistry, Massachusetts Institute of Technology, Cambridge, Massachusetts 02139, USA

[2]Institute for Soldier Nanotechnologies, Massachusetts Institute of Technology, Cambridge, Massachusetts 02139, USA

**ABSTRACT**

We describe a high-speed single-shot multi-frame interferometric imaging technique enabling multiple interferometric images with femtosecond exposure time over a 50 ns event window to be recorded following a single laser-induced excitation event. The stroboscopic illumination of a framing camera is made possible through the use of a doubling cavity which produces a femtosecond pulse train that is synchronized to the gated exposure windows of the individual frames of the camera. The imaging system utilizes a Michelson interferometer to extract phase and ultimately displacement information. We demonstrate the method by monitoring laser-induced deformation and the propagation of high-amplitude acoustic waves in a silicon nitride membrane. The method is applicable to a wide range of fast irreversible phenomena such as crack branching, shock-induced material damage, cavitation and dielectric breakdown.

Topics: high-speed imaging, optical imaging, lasers, Michelson interferometry, multi-frame imaging, microscopy.

[a)] Author to whom correspondence should be addressed: dmitro@mit.edu

# I. INTRODUCTION

Direct optical imaging is a key first step in understanding many complex phenomena. At the micron scale, the imaging of dynamical processes such as fracture[1], propagation of shock waves[2,3], dielectric breakdown[4], and plasma formation[5] require ultra-high-speed diagnostic tools with nanosecond temporal resolution or shorter. Reversible processes may be observed with high temporal resolution utilizing conventional pump-probe measurements with many repetitions, in which a single image is recorded at a variable probe time delay after a pump pulse initiates the event. Irreversible phenomena that may occur in systems that are driven far from equilibrium or into regimes of instability require imaging techniques that can collect multiple images for a single event, i.e. single-shot operation.

Two parameters are crucial in characterizing the ability of single-shot techniques to capture high-speed dynamics: the frame rate ($f$) and the exposure time ($\tau$). On the other hand, $\tau$ dictates the temporal resolution of the imaging system and thus the sharpness of the image. In order to freeze the moment and avoid image blur, the exposure time must be shorter than the time it takes for an object or feature to move by the distance corresponding to the spatial resolution of the optical system. The maximum number of frames $n$ that can be acquired at a frame rate $f$ yields the total window of observation $T$ ($T = n/f$).

Early single-shot imaging was limited to a single image, or multiple exposures on the same image sensor.[6,7] Using light sources ranging from burning magnesium[8] to sparks[9] to flash tubes[10], single images of dynamic events were achieved in the late 19th and early 20th centuries. More recently, ultrafast laser systems have become a powerful tool to obtain single-shot images. The ability of these systems to provide a single fs pulse of light to a camera sensor has allowed for incredibly short exposures times for a single shot. While all of these techniques provided useful information they often lacked the dynamical information provided by multiple-frame systems.

Single-shot multiframe imaging techniques have seen progressive advancement over the last decades. The advent of modern CMOS and CCD cameras with fast readout rates have allowed for nearly continuous recording with microsecond exposure times.[11] On nanosecond timescales the advent of framing cameras with electronic triggering has enabled multiple frames to be collected during a single event (typically with single nanosecond exposure times). At femtosecond and picosecond timescales, spectral encoding techniques have been demonstrated, where the temporal information is encoded into a pulse that has been angled in space to interact with the sample at different time points, or where the pulse has been chirped so that distinct spectral elements encode the temporal information.[12,13] These techniques have exposure times and event windows that are on similar timescales meaning that, for example, the spectrally encoded techniques can achieve fs exposure times (and thus resolve fast dynamics) but can only track these dynamics for femto- to picoseconds.

Currently few techniques are able to collect images with very short exposure times (fs) with long event windows (ns) and high frame rate. Below we describe Multiframe Femtosecond Stroboscopic Imaging (MFSI) and demonstrate the use of a pulse sequence created by an optical cavity synchronized with the frame rate of a high-speed camera to achieve femtosecond time resolved single-shot image sequences. We have previously reported the use of a doubling cavity in combination with a high-speed camera to study complex wave interactions in multi-layered assemblies.[3] Here we extend the technique into interferometric imaging and discuss it in further detail. Crucially this technique provides femtosecond time resolved images (~130 fs) over a tunable long event window (50 ns).

To show the broad utility of this technique, MFSI is employed to monitor in real time, on a single-shot basis, the motion of a silicon nitride membrane following pulsed laser excitation; frames were captured in a Michelson interferometer imaging scheme.[14]

## II. OPTICAL LAYOUT

The optical layout is represented schematically in Fig. 1. A multistage amplified Ti:Sapphire laser (800-nm wavelength) system provided both pump and probe pulses. The stretched oscillator output was amplified at 10 Hz to 4 mJ in a regenerative amplifier, 150 µJ of this was picked off and compressed to 130 fs to act as the pump pulse, and the remainder was amplified further in a 6-pass bowtie amplifier to 20 mJ before being compressed to 130 fs.

The pump light was focused with a lens $L_{pump}$ (30-mm focal length) onto the corner of a free-standing silicon nitride film (100×100 µm, 50-nm thickness) supported on a silicon substrate (purchased from Ted Pella).

The probe light was injected into a frequency-doubling cavity which is formed by two mirrors $M_1$ and $M_2$ that are highly transmissive at 800 nm. $M_1$ is highly reflective at 400 nm, while $M_2$ acts as the "leaking" mirror ($R_{400}$ = 94%) coupling the second harmonic light out of the cavity. Inside the cavity the light was frequency doubled with a β-barium borate (BBO) crystal anti-reflection coated for both 400 nm and 800 nm. The mirrors were positioned 75 cm apart to provide a sequence of pulses separated by 5 ns.[3] The reflectivity of $M_2$ was selected to maximize the energy in the 16$^{th}$ pulse (the maximum number of frames available on our camera), see supplementary material for comparison of different reflectivity of $M_2$.

The light coupled out of the cavity was then directed into an imaging Michelson interferometer. The light was split between reference and sample arms using a polarizing beam splitter (PBS), and the relative intensity of the beams was adjusted using a half-wave plate (HWP) to give the best fringe contrast on the camera. The beams were telescoped using lenses $L_1$ (50-cm focal length) and $L_2$ (3-cm focal length) to yield a beam diameter of ~ 500 µm on the sample surface and to cover the field of view of the camera. In both arms, the polarization of the beams was rotated using quarter wave plates (QWP) and the beams were finally recombined and passed through a polarizer (P). The reference mirror was manually tilted to give interference fringes of equal width on the camera with the desired period (~150 fringes/mm) as described in Veysset et al.[15,16] Interferometric images were then recorded using an ultra-high-speed framing camera (Specialized Imaging, SIMX16). This camera is capable of recording 16 images with as little as 5 ns between frames. By synchronizing the timing between these 16 frames and the incoming pulse train from the cavity, we recorded a series of stroboscopic images with femtosecond temporal resolution over an event window of 50 ns. Each frame has independently controlled pixel gain which allowed the decreasing pulse energy to be compensated (see SI for individual frame camera settings).

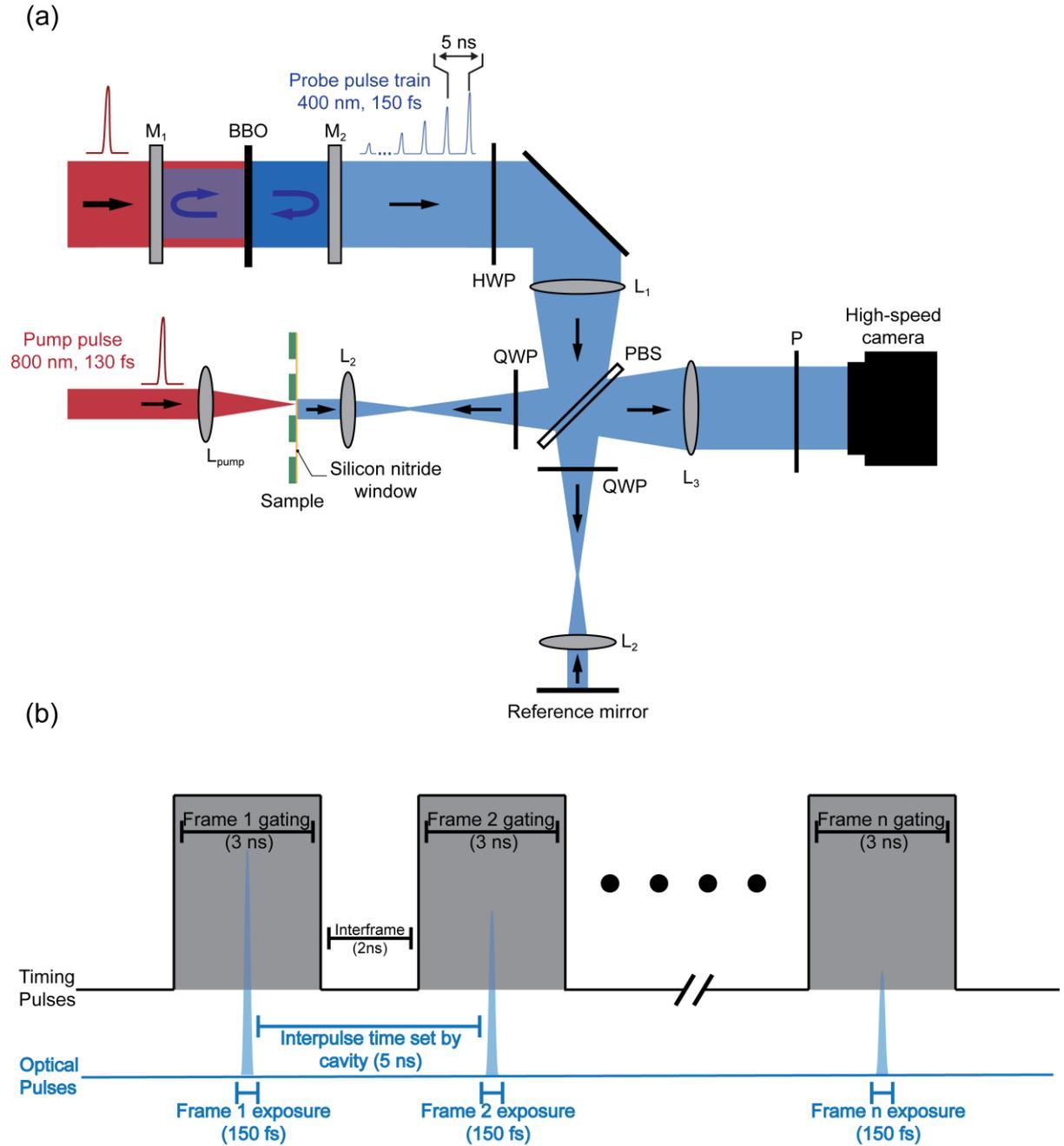

FIG. 1. Schematic illustration of (a) the optical layout and (b) timing scheme for the MFSI technique. The train of pulses coming out of the cavity images the sample surface in a Michelson interferometer configuration. The camera electronic gating is synchronized with the incoming pulse train such that each frame is illuminated with a single fs pulse.

## III. RESULTS AND DISCUSSION

### A. Multiframe imaging

A representative image sequence is shown in Fig. 2. The 130-fs, 150-µJ pump pulse was focused onto the lower right-hand corner of the film, inducing damage and launching an acoustic wave. The images in Fig. 2 show the propagation of this wave with the accompanying deformation of the film. Images were recorded every 5 ns. A triangular dark area grew in throughout the image sequence; this was caused by the extensive bending of the film close to the film's edge, which in turn scattered the imaging light outside of the lateral range of the imaging system. *Post-mortem* examination showed that the film was perforated at the point of laser focus but was intact elsewhere (see Supplementary Material for *post-mortem* confocal microscope images).

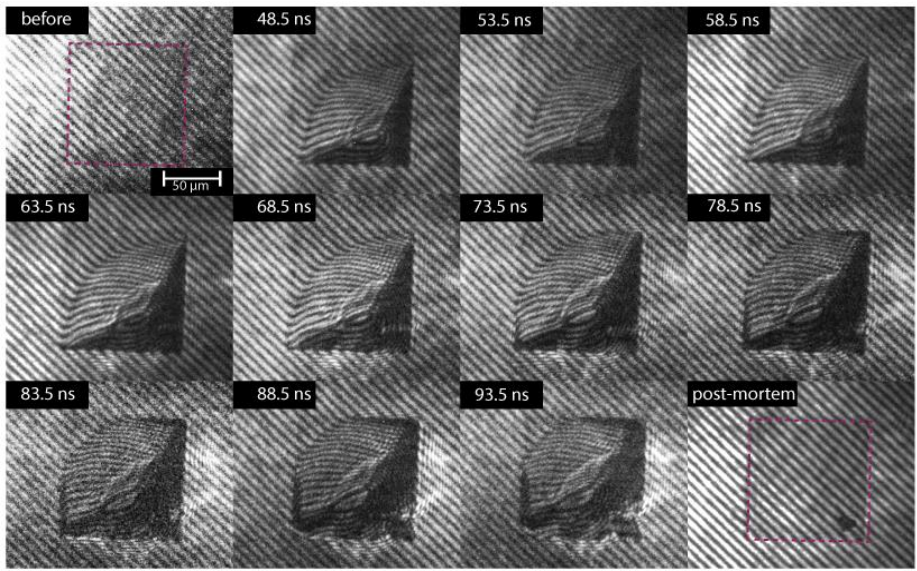

FIG. 2. Single-shot multi-frame images of the evolution of acoustic waves and deformation in a silicon nitride film (film area outlined in "before" and "*post-mortem*" for clarity) following laser excitation. The time stamps at the top of the images indicate the delay between the image acquisition and the excitation. Images are cropped from their original size to show the region of interest (see Supplementary material for full field-of-view image sequence).

**B. Fourier transform profilometry**

The vertical surface elevation field, $u_z(x, y)$, of the silicon nitride membrane can be extracted directly from the interferometric images by analyzing the fringe patterns, containing the phase information, before and after laser excitation. In the present method, the moiré-like pattern was produced using a Michelson interferometer as described in section II. Using a 2D

Fourier transform method[17] with a Hann function for apodization of the spectrum, followed by phase unwrapping (see Supplementary Material), the 2D phase distribution was obtained. This 2D Fourier filtering method assumes slow local variations of illumination intensity and carrier frequency $v_0$. More advanced methods that overcome these limitations have been developed for phase retrieval using, for instance, wavelet transforms and ridge detection methods.[18,19] While these methods for post-acquisition analysis can be implemented on images obtained using MFSI, here we opt for a simple Fourier transform analysis. The phase information was then converted to displacement using the relationship $\Delta\varphi/4\pi = u_Z/\lambda$ where $\Delta\varphi$, $u_Z$, and $\lambda$ denote the phase difference with respect to the reference image, the surface displacement, and the wavelength of the probe beam, respectively.[15,16] The surface displacement for the second frame in Fig. 2 is shown in Fig. 3(a). The wave generated upon laser irradiation of the membrane has propagated on the membrane about 90 microns from the excitation point and has a maximum amplitude of 1.5–2 μm. The wave front profile, extracted from 2D displacement data, is shown for multiple time points in Fig. 3(b) along a radius. The wave travelled at a constant speed of 630 m/s. The wavefront appears mostly circular, although it deviates from this near the edges possibly owing to the attachment of the film to the substrate. In the present images, it is not possible to measure accurately the film displacement close to the excitation spot, as the large slope in the film deflects rays outside of the optical aperture of the imaging setup. After 78.5 ns, the quality of the fringe pattern is degraded (fringe constrast being commensurate with image noise) so the subsequent frames were not analyzed. The resolution of the profilometry with the fringe spacing used in these experiments was 30 nm, corresponding to a minimum phase shift of 0.30 π.

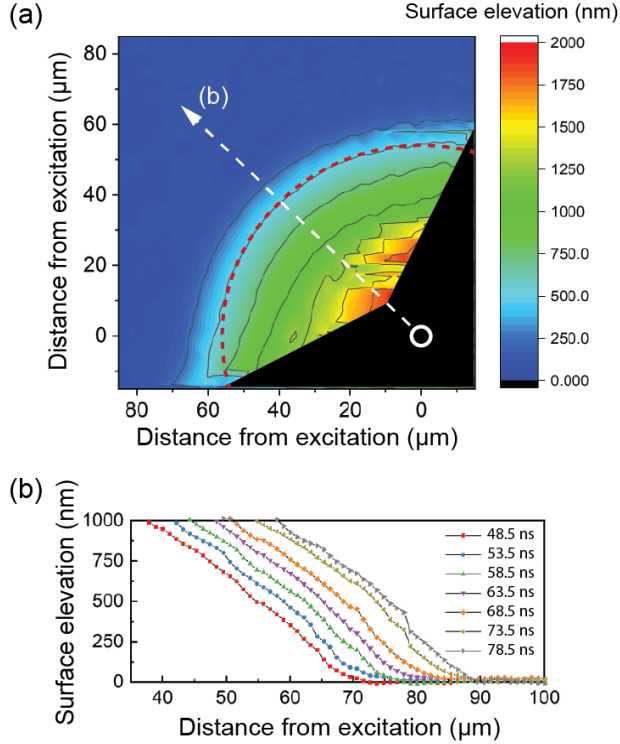

FIG. 3. (a) Surface elevation for frame 2 shown in Fig. 2. Regions where reliable phase extraction was not possible are blacked out. (b) Surface profiles taken along the dotted arrow shown in (a) for multiple frames.

## C. Discussion

At its core the MFSI approach leverages the high temporal resolution of stroboscopic imaging and the multiframe capabilities of modern framing cameras. The resulting technique benefits from the versatility of both of these components: stroboscopic imaging's short time $\tau$ and ability to be interfaced with most imaging techniques, and the framing camera's ability to digitally time multi-frames to high precision, from nanosecond to microsecond interframe times. MFSI benefits from its ability to be adapted into any number of existing imaging techniques, because there is no spectral or temporal shaping of the individual imaging pulses.

Improvements could be realized in the probe pulse train through either higher efficiency cavity design or by the use of a high-power oscillator. A number of commercial manufacturers now produce 80 MHz oscillators with pulse energies sufficient for imaging. With an 80 MHz oscillator providing a never-ending sequence of pulses separated by 12.5 ns, digital control over which pulses were used to record images would allow an essentially limitless event window. As the technique is developed further, one can imagine a combination of MFSI (with or without interferometric imaging) and multi-frame techniques on the fs or ps timescale such as sequentially timed all-optical mapping photography (STAMP)[12] which could yield ultrafast imaging with femtosecond exposure times over 6 decades of event time. In STAMP, the imaging light is chirped and then spectrally separated so that different regions of the image on a single camera sensor correspond to different time delays. Combining this technique with MFSI, using our current multi-frame camera, each CCD would record multiple images separated by fs or ps time intervals, with the sets of images separated by ~ 5 ns, with femtosecond exposure time for all the images, on a single-shot basis.

While we achieved reasonable spatial resolution (2 μm) with a fairly large field of view (480 × 360-μm dimensions), higher spatial resolution could be achieved by replacing the achromatic lens $L_2$ with a high numerical aperture microscope objective.

## IV. CONCLUSIONS

Multiframe Femtosecond Stroboscopic Imaging (MFSI), a technique combining a short exposure time and a long event window, was implemented in an interferometric setup. As a proof of concept, we observed the fast deformation of a laser-excited thin membrane. We demonstrated a short exposure time single-shot multi-frame technique capable of recording a long event window. MFSI is unique in the fact that it allows single-shot imaging in the ($\tau$, $T$) space where imaging techniques remains largely under-developed. Using MFSI coupled with a Michelson interferometer, we tracked the wave speed and deformation properties of a laser-excited silicon nitride film that would be unresolvable without a short exposure time and long event window technique. MFSI is a versatile technique capable of being coupled with numerous imaging techniques and schemes.

## SUPPLEMENTARY MATERIAL

Please see supplementary material for a representative *post-mortem* image of the films, discussion of the cavity design parameters, full-frame unedited images, as well as further discussion of Fourier Transform profilometry.

## ACKNOWLEDGMENTS

DV and AAM would like to thank Felix Hoffman and Sachin Gupta for fruitful discussions and assistance in preliminary testing at early stages of this work. This material is based upon work supported in part by the U. S. Army Research Office through the Institute for Soldier Nanotechnologies, under Cooperative Agreement Number W911NF-18-2-0048.